\documentclass{article}
\usepackage[preprint]{neurips_2026}
\setcitestyle{authoryear,round}

\usepackage[utf8]{inputenc}
\usepackage[T1]{fontenc}
\usepackage{amssymb}
\usepackage{
  hyperref,
  url,
  booktabs,
  amsfonts,
  amsmath,
  nicefrac,
  microtype,
  xcolor,
  graphicx
}
\usepackage{algorithm}
\usepackage[noend]{algpseudocode}
\hypersetup{colorlinks=true, citecolor=blue, linkcolor=blue, urlcolor=blue}

\title{Improving Learned AC-OPF Feasibility under Distribution Shift with PF-in-the-Loop Correction}
\author{
  Afia Afrin \\
  University of Alberta \\
  \texttt{aafrin@ualberta.ca}
  \And
  Omid Ardakanian \\
  University of Alberta \\
  \texttt{oardakan@ualberta.ca}
}
\begin{document}
\maketitle

\begin{abstract}
Learning-based AC-OPF enables fast dispatch, yet its feasibility can degrade under distribution shift. We analyze three state-of-the-art methods and identify a shared deployment-time limitation: none use feedback from the power-flow (PF)-realized state to ensure feasibility of the predicted control.
Using a simple PF-in-the-loop feasibility restoration filter, we empirically show that such feedback substantially improves hard feasibility across four out-of-distribution scenarios with modest runtime overhead.
\end{abstract}
\section{Introduction}
\label{sec:intro}

Learning-based AC optimal power flow (AC-OPF) methods approximate the mapping from system conditions to economical dispatch decisions much faster than conventional iterative solvers. However, their practical value depends on more than just prediction accuracy, which captures dispatch optimality. A near-optimal decision may violate operational limits, especially under distribution shift, where loads, generation availability, or network parameters differ from those seen during training.


Existing methods incorporate physics through different strategies. 
A popular approach is predict-and-reconstruct: 
where a subset of controls is predicted and utilized to recover the remaining ones~\citep{pan2022deepopf,10000707,zamzam2020learning}. 
Physics-informed neural networks embed AC equations, optimality conditions, and physics-derived gradients directly into the architecture or loss function~\citep{nellikkath2022physics,chen2025physics}.  
Other approaches use differentiable correction through equality completion and inequality correction~\citep{donti2021dc3}, or feasibility-seeking procedures reapplied at inference~\citep{nguyen2026fsnet}.
Graph-based and heterogeneous architectures have improved generalization to specific structural shifts, including unseen topologies and contingencies~\citep{gao2023physics,li2025physics,ghamizi2024opf, arowolo2026towards}. 
Complementing this literature, STORM-OPF~\citep{afrin2026storm} showed that learning-based AC-OPF methods with strong nominal performance can suffer severe feasibility degradation under out-of-distribution (OOD) stressed conditions and augmenting training with those stressed samples does not consistently restore feasibility. This finding motivates us to look beyond training-time generalization and examine how predictions are handled after deployment.

Our analysis of three learning-based methods reveals a common phenomenon.
Despite different ways of incorporating physics, all remain \emph{open loop} at deployment:
after producing controls, they do not use violations observed on the PF-realized state to revise the current dispatch. We hypothesize that this feedback is particularly important under distribution shift, where prediction errors can push PF-realized quantities
beyond operating limits. 
To test this hypothesis, we introduce a deliberately simple, model-agnostic \emph{feasibility restoration filter}, that searches over the predicted control vector and evaluates every candidate through nonlinear PF. 

Our main contribution is the empirical finding and its design implication: 
AC-OPF reliability under distribution shift improves substantially when realized constraint violations are fed back to the learned controls. Because PF evaluates fixed controls by solving only the power flow equations, unlike AC-OPF that solves a nonconvex constrained optimization problem over those controls, limited PF feedback can improve OOD feasibility with modest runtime overhead. Thus, we present a practical way to retain the computational advantage of learning-based methods while improving their physical reliability, motivating future architectures that incorporate the control-to-realized-state mapping more directly during training, potentially achieving similar real-world feasibility with fewer PF evaluations.



\section{Methodology}
\label{sec:method}

A learned method may incorporate AC physics during
training or reconstruction, yet in prior work, violations after deployment are not used to revise the dispatch. 
We therefore introduce a
\emph{PF-in-the-loop feasibility restoration layer} that sits between an AC-OPF predictor and the power system. 

\begin{figure}[t]
    \centering
    \includegraphics[trim={100 210 100 230},width=.8\linewidth]{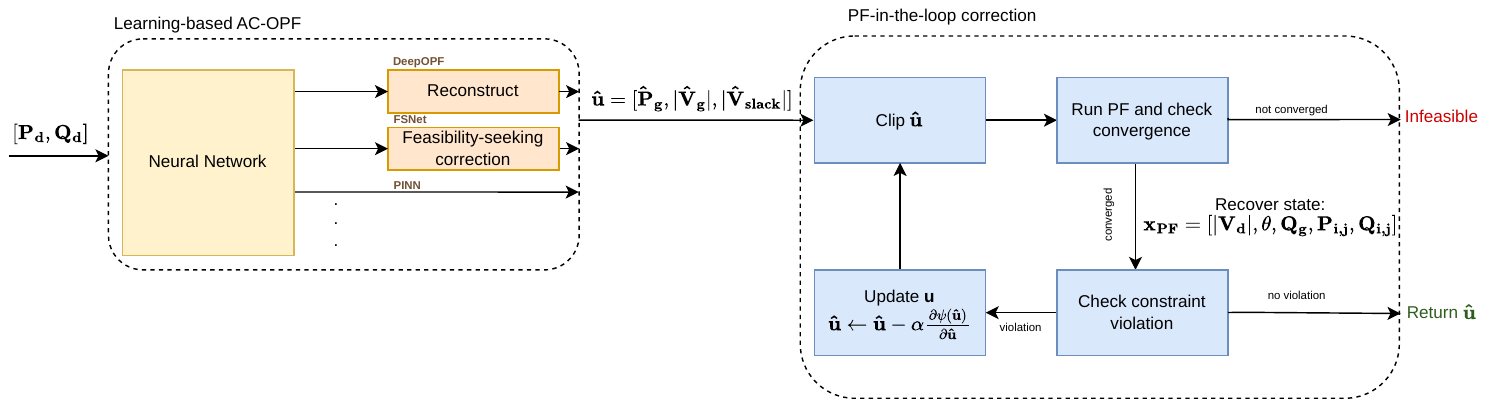}
    \caption{Overview of the proposed correction layer.}
    \label{fig:method}
\end{figure}

As shown in Figure~\ref{fig:method}, the proposed correction layer is model-agnostic. 
A method may predict these controls directly, as in PINN~\citep{nellikkath2022physics}, derive them from a
predicted voltage state, as in FSNet~\citep{nguyen2026fsnet}, use a predict-and-reconstruct method, as in DeepOPF~\citep{pan2022deepopf}, or obtain them through another
approach. The only required interface is the control vector and
the corresponding network instance.

An approximate AC-OPF method produces the control vector, $\mathbf u=\operatorname{col}(\mathbf P_g^{\mathrm{set}},
\mathbf V_g^{\mathrm{set}},V_{\mathrm{slack}}^{\mathrm{set}})$, given the input $(P_d,Q_d)$ representing active and reactive power consumption at load buses. 
Here, $\operatorname{col}$ stacks entries into a column vector,
$\mathbf P_g^{\mathrm{set}}$ contains the active-power
setpoints of generators, and $\mathbf V_g^{\mathrm{set}}$ and
$V_{\mathrm{slack}}^{\mathrm{set}}$ contain the generator- and slack-bus voltage setpoints, respectively.

Given controls
$\mathbf u$
and fixed loads $(P_d,Q_d)$, nonlinear PF solves for the non-slack
voltage angles and PQ-bus voltage magnitudes:
$
\mathbf z = \operatorname{col}
(\boldsymbol\theta_{\mathrm{PV}},\boldsymbol\theta_{\mathrm{PQ}},
\mathbf V_{\mathrm{PQ}}) = \mathrm{PF}(\mathbf u).
$
The complete bus-voltage state combines the PF-solved angles
and magnitudes in $\mathbf z$ with the prescribed voltage
magnitudes in $\mathbf u$: 
\begin{equation}
\mathbf x 
=\operatorname{col}(\boldsymbol\theta,\mathbf V)
=E_z\mathbf z+E_u\mathbf u,
\label{eq:x_pf}
\end{equation}
where the selection matrices $E_z$ and $E_u$ have a $1$ at each position
mapping a solved variable or a prescribed voltage to its bus entry,
and zeros elsewhere. Generator powers and branch currents are calculated from
$\mathbf x$ using the network equations and fixed loads.

Let $V_i$ denote the voltage magnitude of bus~$i$, and $P_{g,i}(\mathbf x)$, $Q_{g,i}(\mathbf x)$, and
$I_{\ell,e}(\mathbf x)$ respectively denote the generator outputs at this bus and
branch currents reconstructed from this voltage state
under fixed loads. 
A converged PF satisfies power balance to solver tolerance but may
violate operating limits. We collect the signed residuals in
$\mathbf r\in\mathbb R^{4|\mathcal G|+2|\mathcal N|+2|\mathcal L|}$ as shown below:
\begin{equation}
\begin{aligned}
\mathbf r
= \operatorname{col}\bigl(&
  (P_{g,i}(\mathbf x)-P_{g,i}^{\max})_{i\in\mathcal G},\\
&
  (P_{g,i}^{\min}-P_{g,i}(\mathbf x))_{i\in\mathcal G},\\
&
  (Q_{g,i}(\mathbf x)-Q_{g,i}^{\max})_{i\in\mathcal G},\\
&
  (Q_{g,i}^{\min}-Q_{g,i}(\mathbf x))_{i\in\mathcal G},\\
&
  (V_i-V_i^{\max})_{i\in\mathcal N},\\
&
  (V_i^{\min}-V_i)_{i\in\mathcal N},\\
&
  (|I_{\ell,e}(\mathbf x)|^2-(I_{\ell,e}^{\max})^2)_{
    (\ell,e)\in\mathcal L\times\{\mathrm f,\mathrm t\}}
\bigr).
\end{aligned}
\label{eq:r}
\end{equation}
Here $\mathcal G$ is the set of generator buses including the slack bus,
$\mathcal N$ is the set of all buses, $\mathcal L$ is the set of all branches, and
$e\in\{\mathrm f,\mathrm t\}$ denotes one end of a branch. $I_{\ell,e}(\mathbf x)$ is the complex current entering branch $\ell$ either at its from-terminal ($e=\mathrm f$) or to-terminal ($e=\mathrm t$). 
All quantities are evaluated at
$\mathbf x$ and expressed in per-unit without further
normalization. Positive residuals indicate violations, and
$\mathbf{r}\preceq\mathbf0$ means that all represented
limits are satisfied.

We guide the correction with
\[
\psi=\left\|[\mathbf r]_+\right\|_2
\] 
where $[\cdot]_+$ takes the componentwise positive part and $\mathbf r$ is the residual calculated at the PF-realized state, $\mathbf x$. 
A lower score need not satisfy every constraint tolerance, so acceptance check uses a separate Boolean test, $\operatorname{HardFeasible}$, on the PF
results. This checks individual generator limits, all bus voltages,
and branch loadings within their specified tolerances.

\paragraph{Obtaining the Gradient $\frac{d\psi}{d\mathbf u}$.} 
As Eq. \ref{eq:x_pf} denotes, $\mathbf u$ determines $\mathbf x$
both explicitly, through $E_u\mathbf u$, and through the PF state
$\mathbf z=\text{PF}(\mathbf u)$, which has no closed form and is defined
only implicitly by the power-balance equations $F(\mathbf z,\mathbf u)=\mathbf 0$.
We therefore obtain $d\psi/d\mathbf u$ with the multivariate chain rule,
taking $d\mathbf z/d\mathbf u$ from the implicit function theorem.
\begin{equation}
\frac{d\psi}{d\mathbf u}
=
\mathbf a\,
\frac{d\mathbf z}{d\mathbf u}
+
\mathbf b,
\qquad
\mathbf a:=\frac{\partial\psi}{\partial\mathbf z},
\quad
\mathbf b:=\frac{\partial\psi}{\partial\mathbf u}.
\label{eq:gradient_chain}
\end{equation}

We use PyTorch autograd to obtain the partial derivatives
$\mathbf a$ and $\mathbf b$.
The remaining term, $d\mathbf z/d\mathbf u$, describes
the sensitivity of the solved PF variables to the controls.
A converged PF satisfies the power-balance mismatch equations
$F(\mathbf z,\mathbf u)=\mathbf0$.
Using multivariate chain rule we have:
\begin{equation}
J\frac{d\mathbf z}{d\mathbf u}+F_{\mathbf u}=\mathbf0,
\qquad
J:=\frac{\partial F}{\partial\mathbf z},
\quad
F_{\mathbf u}:=\frac{\partial F}{\partial\mathbf u}.
\end{equation}
Assuming nonsingular $J$: 
\begin{equation}
\frac{d\mathbf z}{d\mathbf u}=-J^{-1}F_{\mathbf u}.
\end{equation}
Substituting into Eq.~\eqref{eq:gradient_chain} yields
\begin{equation}\nabla \psi = \frac{d\psi}{d\mathbf u}
=-\mathbf aJ^{-1}F_{\mathbf u}+\mathbf b.
\label{eq:gradient_adjoint}
\end{equation}
The Newton--Raphson Jacobian, $J$, and control derivatives, $F_{\mathbf u}$, are computed
analytically using Pandapower's \texttt{dSBus\_dV($\cdot$)} function, which returns the derivatives of the complex power injected at each bus with respect to the bus voltage magnitudes and angles.

The controls mix MW and
per-unit voltages, so we measure changes in scaled controls
$\mathbf v=S^{-1}\mathbf u$, where the diagonal matrix $S$ sets a reference
scale for each control. In our implementation, we use the widest generator active-power range (with a floor
of $1$\,MW) for power controls and $0.1$\,p.u. for voltage controls.

The score $\psi$ is nonsmooth, and its optimal value is known:
$\psi^\star=0$ on any feasible instance. We therefore adopt Polyak's step
size~\citep{polyak1969minimization}, using the known target
$\psi^\star=0$. In scaled coordinates, it takes the shortest step
that drives the local linear model of $\psi$ to zero, adapting the
step length to both the remaining violation and local sensitivity
without step-size tuning:
\[
\Delta\mathbf u
=-\frac{\psi}{\|S\nabla_{\mathbf u}\psi\|_2^2}
S^2\nabla_{\mathbf u}\psi.
\]
Algorithm~\ref{alg:pf_correction} summarizes the proposed feasibility restoration filter. The construction is intentionally simple, as it serves to isolate the value of deployment-time feedback from the
PF-realized state, rather than to claim algorithmic superiority over existing restoration or AC-OPF methods.
Although Polyak's step size has classical convergence guarantees for
convex, possibly nonsmooth, objectives under suitable assumptions, 
these guarantees do not directly extend to our nonconvex violation objective $\psi$. 
We therefore make no general convergence claim and assess the filter empirically through its feasibility recovery and computational cost under a fixed PF budget.

\begin{algorithm}[t]
\small
\caption{PF-in-the-Loop Feasibility Restoration Filter}
\label{alg:pf_correction}
\begin{algorithmic}[1]
\Require Predicted controls $\hat{\mathbf u}$, loads $(P_d,Q_d)$,
control bounds $(\mathbf u^{\min},\mathbf u^{\max})$,
PF budget $B$, control scales
$S=\operatorname{diag}(\mathbf s)$
\Ensure A hard-feasible control vector, or \textsc{Failure}

\State $b\gets0$

\While{$b<B$}
    \State $\hat{\mathbf u}\gets
    \operatorname{clip}
    (\hat{\mathbf u},\mathbf u^{\min},\mathbf u^{\max})$

    \State $\mathbf x_{\mathrm{PF}}
    \gets
    E_{z}(\operatorname{PF}(\hat{\mathbf u};P_d,Q_d)) + E_{u} \hat{\mathbf u}$ 
    \State $b\gets b+1$

    \If{PF does not converge}
        \State \Return \textsc{Failure} \Comment{counted as infeasible}
    \EndIf

    \If{$\operatorname{HardFeasible}(\mathbf x_{\mathrm{PF}})$}
        \State \Return $\hat{\mathbf u}$
    \EndIf

    \If{$b\geq B$}
        \State \textbf{break}
    \EndIf

    \State get $\mathbf r$ using Eq \eqref{eq:r}
    \State {$\psi=\left\|[\mathbf r]_+\right\|_2$}

    \State $\nabla\psi
    \gets$ Eq.~\eqref{eq:gradient_adjoint} at $\mathbf x_{\mathrm{PF}}$
    \Comment{no additional PF evaluation}

    \State $\hat{\mathbf u}\gets
    \hat{\mathbf u}
    -
    \dfrac{\psi}{\|S\nabla\psi\|_2^2}\,S^2\nabla\psi$
    \Comment{Polyak step, $\psi^\star=0$}
\EndWhile

\State \Return \textsc{Failure}
\end{algorithmic}
\end{algorithm}

\vspace{-2mm}

\section{Experimental Setup}
\label{sec:setup}

We evaluate three representative AC-OPF solvers, namely DeepOPF, PINN, and FSNet, on IEEE-14 bus system, representing a small approximation of the American electric power grid as of February 1962. 
Each model is trained only on nominal samples and evaluated on five scenarios, each comprising 1,000 instances generated using distinct load perturbations.
The in-distribution scenarios are generated by independently perturbing loads within $[0.9,1.1]$ of their base values. 
The four OOD cases represent test-time conditions in the same power system, such as extreme weather, light load, geographic load redistribution, and heavy load, following their definition in~\cite{afrin2026storm}.

We use Pandapower's~\citep{pandapower.2018} default iterative solver (PDIPM) as the reference solver. 
As nonconvex AC-OPF may converge to different local
solutions, a multi-start version of PDIPM, PDIPM-MS, runs PDIPM from five initializations and retains the
cheapest feasible result. PDIPM-MS is used only as a stronger optimality reference; its higher runtime is not used for comparison.

The filter is invoked only when the predicted dispatch fails the hard-feasibility check. We use a correction budget of \(B=40\) PF evaluations in our implementation, i.e., up to 40 iterations, since each iteration costs one PF evaluation. The bound vectors \(\mathbf u^{\min}\) and \(\mathbf u^{\max}\) comprise the generator active-power limits and the PV- and slack-bus voltage limits specified by each evaluated scenario.

\subsection{Evaluation Metrics}
\label{sec:metrics}

We use the same evaluation metrics as
STORM-OPF~\citep{afrin2026storm}, namely feasibility, operating cost, and runtime, summarized below for completeness.

\paragraph{Hard feasibility rate.}
For every test instance, we apply the predicted or corrected controls and run
nonlinear AC PF. A dispatch is feasible only if PF converges and all applicable
operating constraints are satisfied: active- and reactive-power limits of
all generators, voltage-magnitude limits at PQ
buses, and loading limits of in-service lines and transformers. We allow
numerical tolerances of $10^{-3}$ p.u.\ for voltage, $10^{-3}\%$ for loading, and the larger of $10^{-6}$ MW/Mvar or $0.1\%$ of the
corresponding generator bound. For any method $m$, evaluated on $N_{\mathrm{test}}$ samples, the feasibility rate is
$
  F^m
  =
  \frac{100}{N_{\mathrm{test}}}
  \sum_{k=1}^{N_{\mathrm{test}}}
  \mathbb{I}_k^m,
$
where $\mathbb{I}_k^m=1$ only when instance $k$ passes the  hard feasibility test.

\paragraph{Operating cost.}
Operating cost is evaluated only for feasible dispatches. For each
instance, we sum the polynomial generation costs stored in the network and
report the ratio to optimization references:
$
  r_k^m
  =
  \frac{C_k^m}{C_k^{\mathrm{ref}}}.
$
The reported value is the mean of the per-instance ratios over samples on which both method $m$ and the reference are feasible. 
Because methods can be feasible on different subsets,
cost ratios from columns with different feasible sets are not direct paired
before-and-after comparisons.

\paragraph{Runtime.}
For learned pipelines, runtime is measured per sample from an available
network instance to a verified feasibility verdict, including model inference,
the proposed correction when activated, and PF verification. We report the
mean runtime for each baseline. Single-start
PDIPM reports its solve time. For multi-start PDIPM, the reported runtime is the sum of runtimes of all individual starts. 

\section{Results}
\label{sec:results}

\begin{table*}[t]
  \caption{Hard AC feasibility (\%), cost relative to the optimization
  references, and runtime. ``Native'' is the original predictor; ``Corr.''
  applies the PF-in-the-loop filter with a 40-PF budget.}
  \label{tab:main}
  \centering
  \scriptsize
  \setlength{\tabcolsep}{3.2pt}
  \resizebox{\textwidth}{!}{%
  \begin{tabular}{@{}lcccccccc@{}}
    \toprule
    & \multicolumn{2}{c}{DeepOPF}
    & \multicolumn{2}{c}{PINN}
    & \multicolumn{2}{c}{FSNet}
    & \multicolumn{2}{c}{Optimization references}\\
    \cmidrule(lr){2-3}
    \cmidrule(lr){4-5}
    \cmidrule(lr){6-7}
    \cmidrule(l){8-9}
    Scenario / Eval. Metric
    & Native & Corr.
    & Native & Corr.
    & Native & Corr.
    & PDIPM & PDIPM-MS\\
    \midrule
    In-distribution
    & 100.0 & 100.0
    & 99.6 & 100.0
    & 100.0 & 100.0
    & 100.0 & 100.0\\
    \midrule
    Extreme weather
    & 0.0 & 100.0
    & 0.0 & 100.0
    & 100.0 & 100.0
    & 100.0 & 100.0\\
    Light load
    & 0.0 & 100.0
    & 0.0 & 100.0
    & 94.6 & 100.0
    & 100.0 & 100.0\\
    Geographic shift
    & 50.9 & 100.0
    & 7.5 & 100.0
    & 99.5 & 100.0
    & 100.0 & 100.0\\
    Heavy load
    & 79.2 & 100.0
    & 0.0 & 100.0
    & 99.9 & 100.0
    & 100.0 & 100.0\\
    \midrule
    Mean feasibility
    & 46.0 & 100.0
    & 21.4 & 100.0
    & 98.8 & 100.0
    & 100.0 & 100.0\\
    Cost normalized by PDIPM cost
    & 1.02 & 0.96
    & 1.00 & 0.94
    & 1.02 & 1.02
    & 1.00 & 0.93\\
    Cost normalized by PDIPM-MS cost
    & 1.02 & 1.04
    & 1.00 & 1.01
    & 1.10 & 1.10
    & 1.13 & 1.00\\
    Runtime (ms/sample)$\downarrow$
    & 32.8 & 69.6
    & 29.0 & 67.8
    & 63.7 & 64.2
    & 219.7 & 3410.9\\
    \bottomrule
\end{tabular}
  }
\end{table*}

Table~\ref{tab:main} shows substantial feasibility improvements for all three
architectures across all OOD scenarios. Mean feasibility rate increases by $54.0\%$, $79.9\%$, and $0.1\%$ for DeepOPF, PINN, and FSNet, respectively. All corrected pipelines remain faster than the 219.7\,ms single-start PDIPM
reference on case14, at around 33--70\,ms per sample. 

\section{Limitations}
\label{sec:limitations}

The experiments use one test network, four constructed shifts, and selected model checkpoints. Each iteration costs one PF evaluation and one sparse Jacobian solve regardless of the control dimension, but scalability to larger systems remains to be verified.
The local search can encounter non-convergent PF candidates and offers
no guarantee of finding a feasible point within its budget.
The filter also ignores generation cost while searching for a feasible dispatch. Taken together, these limitations leave open several questions, warranting further study on incorporating cost-preservation, evaluating larger networks and topology changes, and studying
fallback policies for samples that exhaust the correction budget.

\section{Conclusion}
\label{sec:conclusion}



We investigate a common source of learned AC-OPF infeasibility under distribution shift. Our analytical finding suggests, existing methods remain open loop at deployment, as violations in the PF-realized state are not used to revise the predicted control dispatch. This gap becomes especially consequential when operating conditions differ from the training distribution.
Using a simple, training-free PF-in-the-loop correction, we show across three predictor families that realized-state feedback substantially improves feasibility. We note that the proposed feasibility restoration filter is intended as a design probe, rather than as a new AC-OPF solver or standalone algorithmic contribution.



Future work should integrate this feedback more directly into model training, explore the tradeoff between runtime and feasibility rate governed by the budget, 
preserve dispatch
cost, and combine bounded correction with a reliable optimization fallback.

\bibliographystyle{plainnat}
\bibliography{references}

\clearpage

\end{document}